\documentclass{optica-article}

\journal{opticajournal} 

\articletype{Research Article}

\usepackage{lineno}
\usepackage{bbm}
\usepackage{color}
\usepackage{comment}

\begin{document}

\title{Observation of quantum nonlocality in Greenberger-Horne-Zeilinger entanglement on a silicon chip}

\author{Leizhen Chen,\authormark{1,5}
        Bochi Wu,\authormark{1,5}
        Liangliang Lu,\authormark{2}
        Kai Wang,\authormark{1}
        Yanqing Lu,\authormark{1}
        Shining Zhu,\authormark{1}
        Xiao-Song Ma,\authormark{1,3,4,*}}

\address{\authormark{1}National Laboratory of Solid-state Microstructures, School of Physics, College of Engineering and Applied Sciences, Collaborative Innovation Center of Advanced Microstructures, Nanjing University, Nanjing 210093, China\\
\authormark{2}Key Laboratory of Optoelectronic Technology of Jiangsu Province, School of Physical Science and Technology, Nanjing Normal University, Nanjing 210023, China\\
\authormark{3}Synergetic Innovation Center of Quantum Information and Quantum Physics, University of Science and Technology of China, Hefei, Anhui 230026, China\\
\authormark{4}Hefei National Laboratory, Hefei 230088, China\\
\authormark{5}These authors contributed equally: $\rm{Leizhen\ Chen, Bochi\ Wu}$}

\email{\authormark{*}Xiaosong.Ma@nju.edu.cn} 

\begin{abstract*} 
Nonlocality is the defining feature of quantum entanglement. Entangled states with multiple particles are of crucial importance in fundamental tests of quantum physics as well as in many quantum information tasks. One of the archetypal multipartite quantum states, Greenberger-Horne-Zeilinger (GHZ) state, allows one to observe the striking conflict of quantum physics to local realism in the so-called all-versus-nothing way. This is profoundly different from Bell’s theorem for two particles, which relies on statistical predictions. Here, we demonstrate an integrated photonic chip capable of generating and manipulating the four-photon GHZ state. We perform a complete characterization of the four-photon GHZ state using quantum state tomography and obtain a state fidelity of 0.729$\pm$0.006. We further use the all-versus-nothing test and the Mermin inequalities to witness the quantum nonlocality of GHZ entanglement. Our work paves the way to perform fundamental tests of quantum physics with complex integrated quantum devices.

\end{abstract*}

\section{Introduction}
{Entanglement lies at the heart of quantum mechanics and is an important resource for quantum information processing. This stems from the non-classical features of entangled states. The statistics of certain multipartite quantum systems are incompatible with the predictions of local realism \cite{bell1964einstein}, and therefore are considered to be nonlocal \cite{RevModPhys.86.419}. }
To understand and prove this property of entanglement, much ingenious work has been done in different quantum systems to close the loopholes in the bipartite Bell experiments \cite{PhysRevD.10.526,PhysRevLett.49.1804,PhysRevLett.81.5039,rowe2001experimental,PhysRevLett.100.150404,scheidl2010violation,giustina2013bell,PhysRevLett.111.130406,hensen2015loophole,PhysRevLett.115.250401,PhysRevLett.115.250402,PhysRevLett.119.010402,PhysRevLett.121.080404}. As the number of particles and the dimensions of the entangled states increase, nonlocality shows a much richer and more complex structure \cite{RevModPhys.84.777,erhard2020advances}. In their seminal works, Greenberger, Horne and Zeilinger (GHZ) discovered that a certain class of maximally-entangled three-particle state (GHZ state) does not require probabilistic Bell-like inequalities to refute the local realism model \cite{greenberger1989going,greenberger1990bell}. Since this discovery, the GHZ state has become one of the most studied multi-particle entangled state in various physical systems, including bulk optics \cite{PhysRevLett.82.1345,pan2000experimental,PhysRevLett.86.4435,PhysRevLett.91.180401,lu2007experimental,erven2014experimental,PhysRevLett.115.260402,tsujimoto2018high}, integrated photonics \cite{adcock2019programmable,llewellyn2020chip,pont2022high}, superconducting qubits \cite{dicarlo2010preparation,kelly2015state,PhysRevLett.119.180511,cao2023generation}, trapped ions and atoms \cite{PhysRevLett.106.130506,PhysRevLett.112.100403,zhao2021creation} and solid-state qubits \cite{pompili2021realization}. Recently, GHZ states are moving toward more particles \cite{PhysRevLett.117.210502,PhysRevLett.120.260502,omran2019generation} and higher dimensions \cite{erhard2018experimental,bao2023very}, paving the way for large-scale quantum computation and multiparty quantum communication.

Silicon-based integrated photonic circuits promise desirable properties for photonic quantum technology. It offers dense device integration, high optical nonlinearity, and robust phase stability. Some recent works have demonstrated the ability to generate complex and programmable multi-photon quantum states with integrated optics \cite{vigliar2021error,PhysRevLett.130.223601,bao2023very,zheng2023multichip}. 
To study the fundamental quantum physics in more complex systems, it is crucial to realize high-fidelity operations in photonic circuitries with increasing complexity, and a larger number of photons with high dimensions. However, few of the previous demonstrations studied the quantum nonlocality of GHZ states with integrated photonics, such as all-versus-nothing nonlocal trait and Mermin conflict between quantum mechanics and local realism. They are mainly based on bulk optical setups with polarization-entangled photons \cite{pan2000experimental,PhysRevLett.91.180401,PhysRevLett.115.260402} or photons with orbital-angular-momentum modes \cite{erhard2018experimental}. 
It is therefore of crucial importance to investigate whether quantum nonlocality can be produced and verified with the scalable quantum photonic chips. In this work, we go beyond the on-chip Bell experiments with two photons, and generate and employ the three- and four-photon GHZ states to observe quantum nonlocality.
We use a fully-integrated silicon chip, including path-entangled quantum light modules with advanced dual-Mach-Zehnder interferometer micro-ring sources \cite{tison2017path,vernon2017truly,lu2020three}, wavelength routing, quantum state parity sorter as well as single-photon tomographic modules, to generate, manipulate, and topographically reconstruct the GHZ states. Moreover, we measure the quantum nonlocality of GHZ entanglement via both the all-versus-nothing (AVN) test \cite{greenberger1990bell,PhysRevLett.91.180401} and the Mermin inequalities \cite{PhysRevLett.65.1838,PhysRevA.46.5375}, showing the striking conflict of quantum physics to local realism in the multiphoton scenario. Photons are encoded along the path mode of waveguides, which can be flexibly extended to high dimensions and  potentially useful for future large-scale quantum information processing and quantum networks. It should be noted that a recent work by M. Pont et al. \cite{pont2022high} has reported a powerful but different platform of a solid-state quantum-dot photon source and glass-based photonic circuits to generate high-quality GHZ entanglement and certify its quantum nonlocality via a Bell-like inequality.

\section{Experimental setup}
\begin{figure*}[tbp]
    \centering
	\includegraphics[width=1\textwidth]{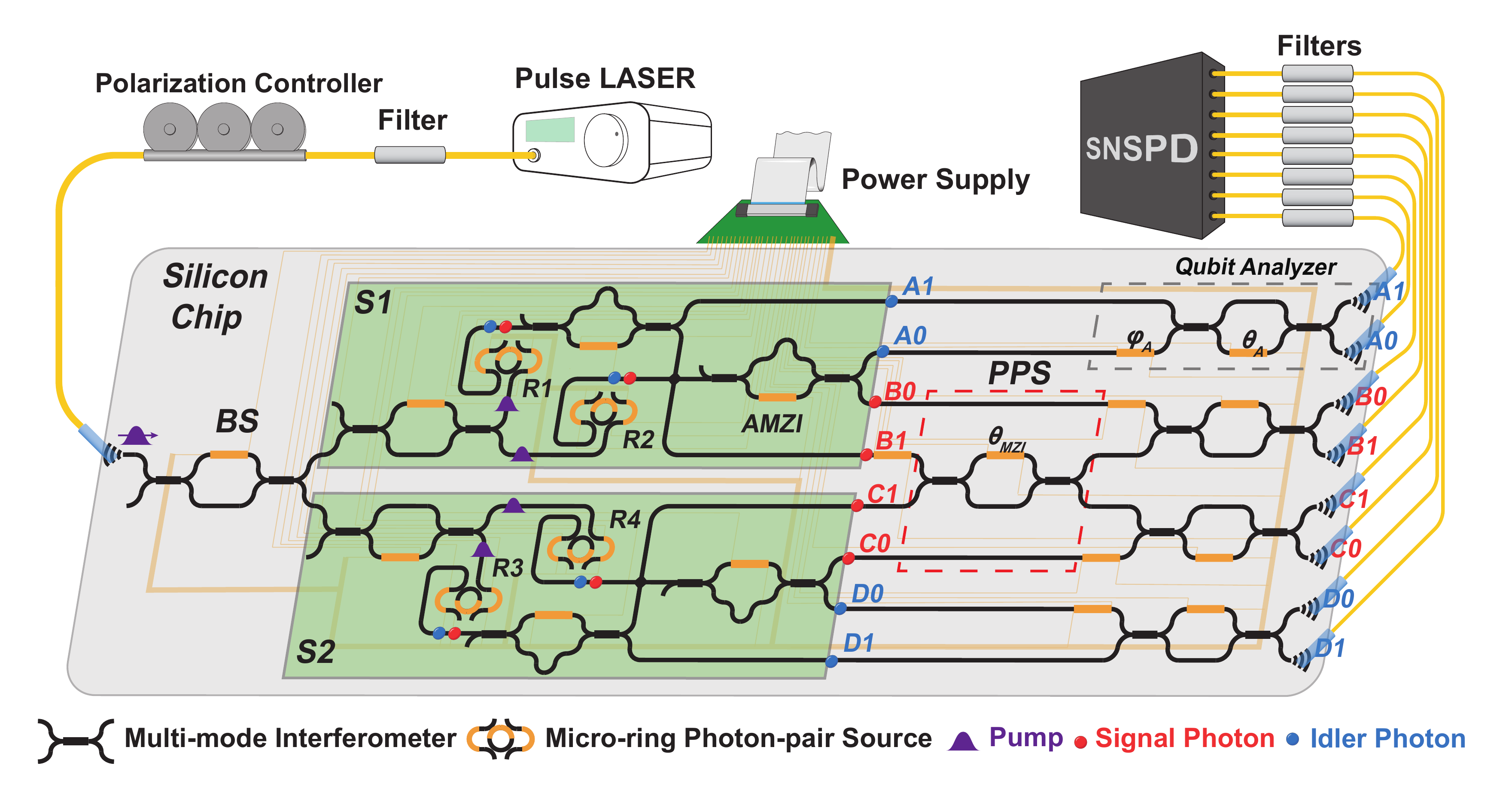}
	\caption{\label{fig1} Experimental setup for generating and manipulating the four-photon GHZ state. A pulsed laser is filtered, polarized and grating-coupled into the chip. Then, the pulse coherently pumps two path-entangled photon-pair source modules, S1 and S2, which emit path-entangled photon pairs with different wavelengths (1544.5 nm and 1556.5 nm). In each source module, two micro-rings generate photon-pairs, and asymmetric Mach–Zehnder interferometers (AMZIs) separate photons. Then, on-chip path-mode parity sorter (PPS) is used to combine and sort the signal photons B and C according to their path modes. After that, photons are distributed to four qubit analyzers, coupled out, filtered, and finally detected by superconducting nanowire single-photon detectors (SNSPDs). Simultaneous detection of one photon in each outport heralds the generation of four-photon GHZ state. All on-chip phase shifters and micro-rings are controlled with a programmable multi-channel power supply.
    }     
\end{figure*}
The integrated photonic circuit consists of two path-entangled photon-pair source modules (S1 and S2), an on-chip path-mode parity sorter (PPS), and four qubit analyzers [Fig.\ref{fig1}]. A pulsed laser is grating-coupled into the chip and split with a beam spliter (BS). Then, S1 and S2 are coherently pumped by the laser, and each emits entangled photon pairs in the Bell state,
\begin{equation}
    \left|\Psi_2\right>=1/\sqrt{2}\left(\left|00\right>+\left|11\right>\right).
    \label{eq1}
\end{equation}
Note that $\left|0\right>$ and $\left|1\right>$ stand for different path modes. 
In each path-entangled photon-pair source module, we use two integrated micro-ring photon-pair sources to emit path-encoded photon pairs and asymmetric Mach–Zehnder interferometers (AMZIs) to separate them according to different wavelengths of signal and idler photons. Hence, there are four micro-rings on the chip and they are all coherently pumped to provide photon pairs for producing the four-photon GHZ state. Then, a Mach–Zehnder interferometer (MZI) and two waveguides constitute a controllable PPS to combine and interfere photon B and C. After evolution of photons at the PPS, the simultaneous detection of photons in each output A-D gives the output four-photon GHZ state (see Appendix A for a detailed description on GHZ-entanglement generation),
\begin{equation}
    \left|\Psi_4\right>=1/\sqrt{2}\left(\left|0000\right>+\left|1111\right>\right).
    \label{eq:ghz}
\end{equation}
Note that $\left|0000\right>$ and $\left|1111\right>$ stand for $\left|0_{\rm{A}}\right>\otimes\left|0_{\rm{B}}\right>\otimes\left|0_{\rm{C}}\right>\otimes\left|0_{\rm{D}}\right>$ and $\left|1_{\rm{A}}\right>\otimes\left|1_{\rm{B}}\right>\otimes\left|1_{\rm{C}}\right>\otimes\left|1_{\rm{D}}\right>$, respectively. The post-selection probability to observe the GHZ state is 0.25 from all events of four-photon generation including the possibility that one source module S1 (or S2) generates two pairs of photons. 
Then, four path-mode qubit analyzers with MZIs and phase shifters are used to measure the post-selected multi-photon states. After that, all photons are coupled out with grating couplers, filtered with off-chip filters, and detected with superconducting nanowire single-photon detectors (SNSPDs). The single-photon detection events are recorded by an FPGA-based coincidence logic unit which further calculates and outputs both two- and four-fold coincidence counts between different path modes.

\section{Results}
The silicon photonic chip was manufactured at the Advanced Micro Foundry (AMF), with 500 nm$\times$220 nm fully-etched SOI waveguides and TiN heaters placed 2 µm above waveguides. We use a picosecond pulsed pump laser of $\sim$ 4 mW power, 60 MHz repetition rate, and 1550.5 nm wavelength with $\sim$ 0.8 nm bandwidth. The laser is coupled into the chip by grating couplers with a coupling loss of $\sim5$ dB/facet. After the laser propagating on chip and splitting with MZIs, four micro-ring sources (R1, R2 in S1 and R3, R4 in S2) are coherently pumped with $\sim0.2$ mW power on each micro-ring. 
Then, photon pairs with different wavelengths (1544.5 nm and 1556.5 nm) are generated via spontaneous four wave mixing (SFWM) process and routed according to their wavelengths with AMZIs. The measured pair-generation rate of each micro-ring ranges from 1400 Hz to 2400 Hz, and the four-fold coincidence count rate for the four-photon GHZ state is $380\pm32$ per hour. 
\begin{figure}[tbp]
    \centering
	\includegraphics[width=0.85\textwidth]{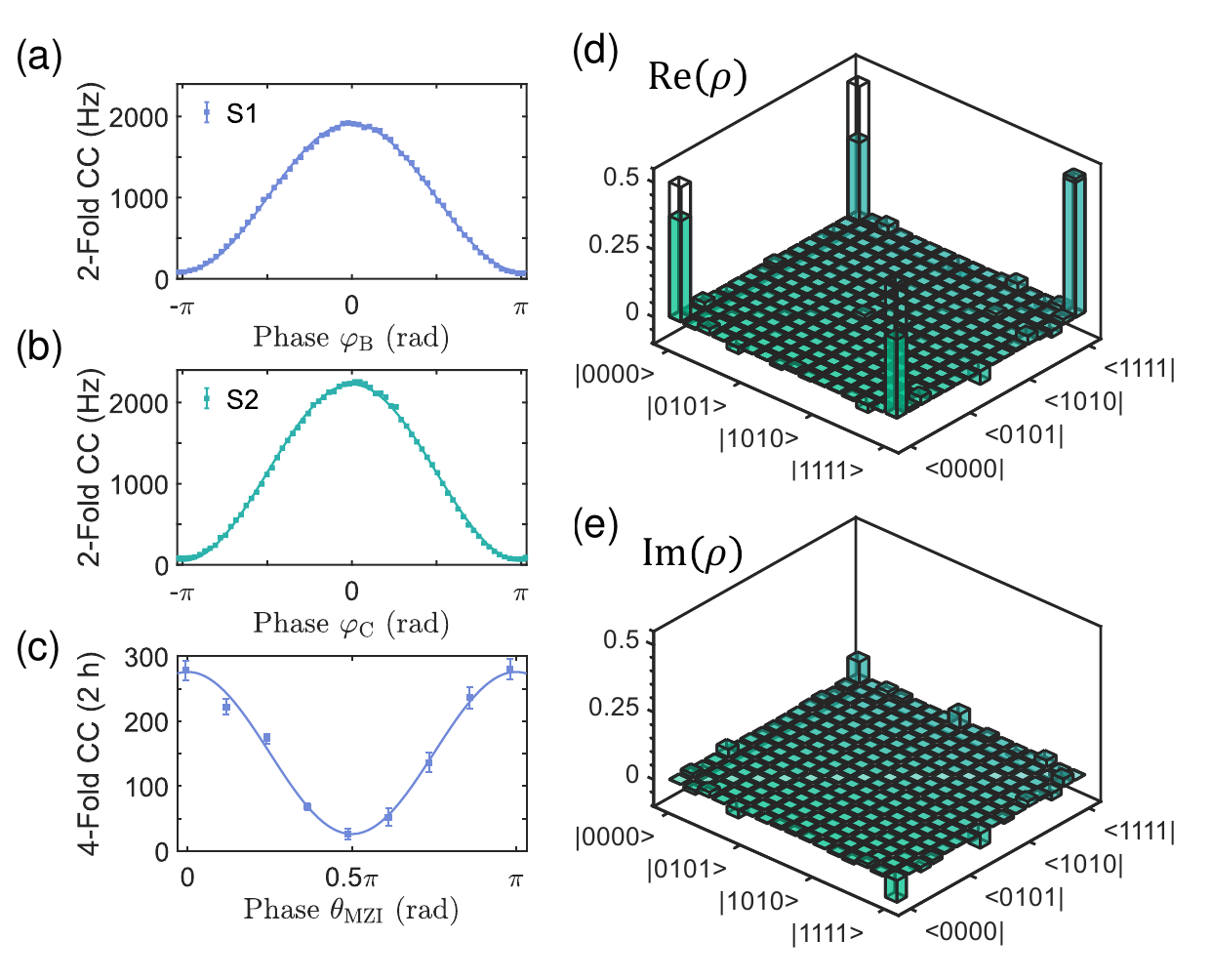}
	\caption{\label{fig2} Experimental results of four-photon GHZ state. (a), (b) Two-photon interference fringes of source modules S1 and S2, indicating high-quality path correlation of qubit entanglement. The measured path-correlation visibilities are $0.935\pm0.004$ and $0.938\pm0.003$, respectively. (c) Heralded-HOM interference between micro-rings R1 and R3, showing indistinguishability of signal photons from S1 and S2. The heralded-HOM visibility is $0.814\pm 0.038$. (d), (e) Real and imaginary part of the density matrices of four-photon GHZ states. Experimental results are shown with colored bars. Theoretical predictions are shown with the wire grid. The quantum-state fidelity is $0.729\pm0.006$. Uncertainties are obtained from Monte Carlo simulations with Poisson counting statistics.}
\end{figure}

First, we verify the fidelity of the two-photon Bell state $\left|\Psi_2\right>$ generated from each path-entangled photon-pair source module (S1 or S2) by measuring the path-correlation in mutually unbiased bases (MUBs) \cite{silverstone2015qubit}. We project idler photons into the basis $1/\sqrt{2}\left(\left|0\right>+\left|1\right>\right)$ and signal photons the basis $1/\sqrt{2}\left(\left|0\right>+e^{i\phi}\left|1\right>\right)$. By scanning the phase $\phi$, two-photon interference fringes are obtained from the two-fold coincidence between signal and idler photons [Fig.\ref{fig2}(a) and (b)].
{Here, the visibility of path-correlation fringes is defined as $V_{path-cor.}=(CC_{max}-CC_{min})/(CC_{max}+CC_{min})$, where $CC_{max}$ and $CC_{min}$ are the maximum and the minimum of coincidence counts.}
 We obtain path-correlation visibilities of $0.935\pm0.004$ for S1 and $0.938\pm0.003$ for S2, respectively. The deviation of visibilities is due to multi-pair generation events and the spectral distinguishability of photons from different micro-rings. From the relationship between fidelity and visibility in a Werner state \cite{zhang2017simultaneous}, $F=(1+3V)/4$, the estimated fidelities for two-photon entangled states are $0.952\pm0.002$ for S1 and $0.953\pm0.002$ for S2, indicating high quality of qubit-entanglement of our sources. 

Next, we investigate the indistinguishability between source modules S1 and S2 via four-photon heralded Hong-Ou-Mandel (HOM) interference \cite{faruque2018chip}. In this case, signal photons from S1 and S2 are combined and interfered by a tunable MZI in the PPS. By scanning the splitting ratio of the MZI, we obtain a visibility of $0.814\pm 0.038$ from the interference fringe [Fig.\ref{fig2}(c)]. Here, the visibility of heralded-HOM fringe is defined as $V_{HHOM}=\left(CC_{max}/2-CC_{min}\right)/\left(CC_{max}/2\right)$ (see Appendix C for details). 
The reduced visibilities are mainly due to the large accidentals from multi-pair generation events and the spectral impurity of micro-ring photon sources.

We then proceed to characterize the GHZ state with complete quantum-state tomography (QST) \cite{PhysRevA.64.052312}. The density matrices of the four-photon GHZ state reconstructed from the complete set of 81 measurements are displayed in Fig.\ref{fig2}(d) and (e). The quantum-state fidelity with ideal case is $F_{GHZ_4}=\text{Tr}\left(\rho_{\text{exp}}\rho_{\text{ideal}}\right)=0.729\pm0.006$, providing quantitative characterization of GHZ states. Also, a two-setting witness measurement gives $\left<W_{GHZ_4}\right>=-0.383\pm0.042$, showing the genuine four-photon entanglement. 
Note that the entanglement witness is defined by $W_{GHZ_4}:=3\mathbbm{1}-2\left[(S_1+\mathbbm{1})/{2}+\prod_{i=2}^4(S^{(i)}_2+\mathbbm{1})/2\right]$ \cite{PhysRevLett.94.060501}, where two measurement settings are $S_1=\sigma_x^{(1)}\sigma_x^{(2)}\sigma_x^{(3)}\sigma_x^{(4)}$ and $S_2^{(i)}=\sigma_z^{(i-1)}\sigma_z^{(i)}$. 
By performing single-qubit projective measurement in $1/\sqrt{2}\left(\left|0\right>+\left|1\right>\right)$, we obtain three-photon GHZ states, $\left|\Psi_3\right>=1/\sqrt{2}\left(\left|000\right>+\left|111\right>\right)$, shared between the remaining photons. The average fidelity of three-photon GHZ states is $0.742\pm0.004$ [Tab.\ref{tab1}].

Having established the high-quality GHZ entanglement, we first demonstrate its quantum nonlocality via the AVN test. The AVN test is based on perfect correlations in multipartite systems and offers a logical contradiction between local-hidden-variable models and quantum mechanics directly without any inequality \cite{greenberger1990bell,pan2000experimental,PhysRevLett.91.180401}. 
The AVN test for 4-photon GHZ state is based on the following equations:
\begin{eqnarray}
XXXX &=& 1, \label{AVN1}\\
YYYY &=& 1, \label{AVN2}\\
XXYY &=& -1, \label{AVN3}\\
XYXY &=& -1, \label{AVN4}\\
XYYX &=& -1, \label{AVN5}\\ 
YXXY &=& -1, \label{AVN6}\\
YXYX &=& -1, \label{AVN7}\\ 
YYXX &=& -1. \label{AVN8}
\end{eqnarray}
\begin{figure}[bp]
    \centering
	\includegraphics[width=0.5\textwidth]{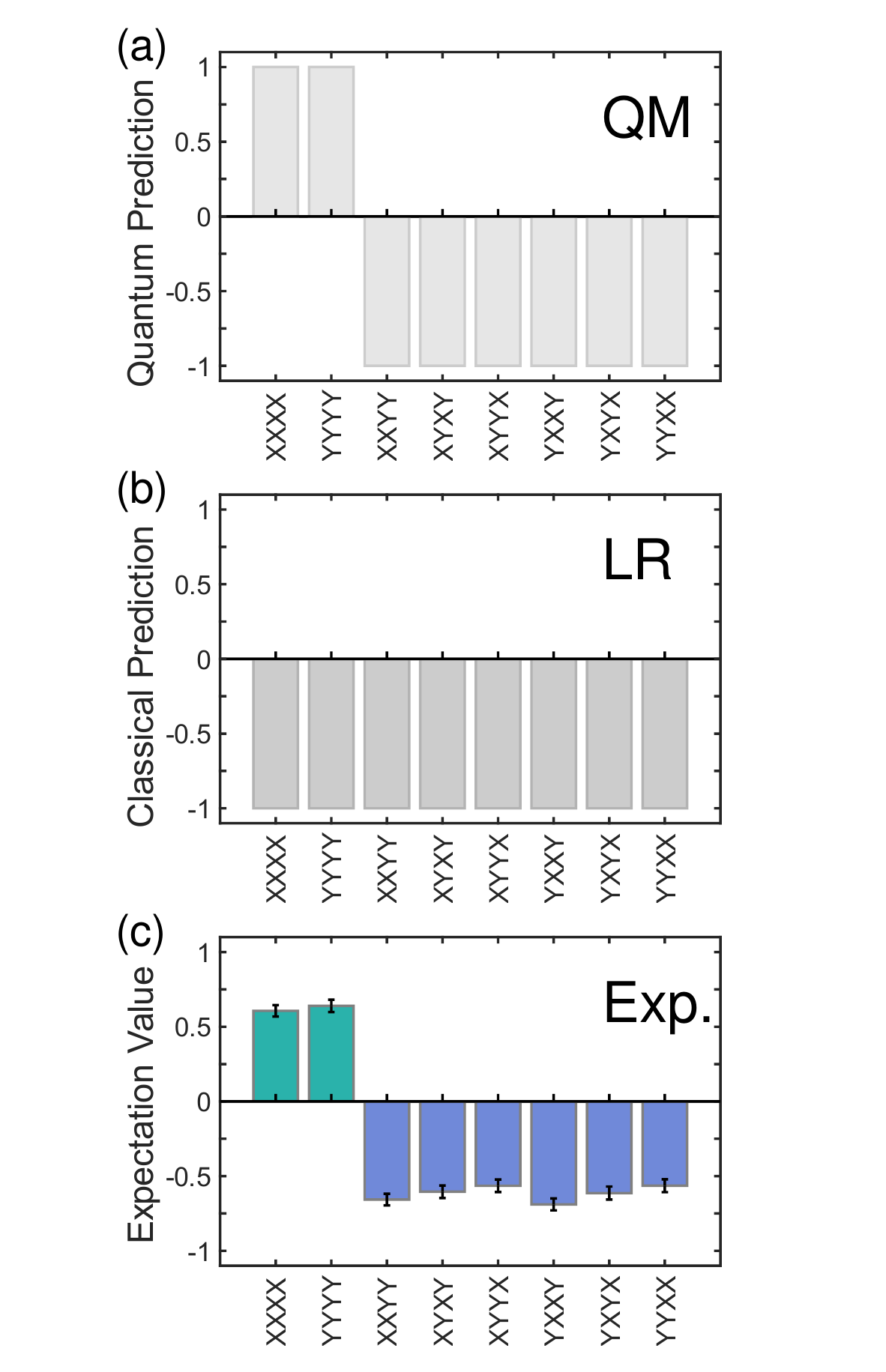}
	\caption{\label{fig3}All-versus-nothing test for four-photon GHZ states. (a) Predictions of quantum mechanics are based on perfect correlations of GHZ states. (b) Local realism contradict quantum mechanics on at least two predictions. (c) Experimental results are in agreement with the quantum mechanics and in conflict with the local realism.}
\end{figure}
Note that $X_{\rm{i}}$ ($Y_{\rm{i}}$) denotes the value, +1 or -1, obtained when measuring $\sigma_x^{(i)}$ ($\sigma_y^{(i)}$) on photon i. Quantum mechanics predicts that all equations can be satisfied, while local realistic theories can reproduce only 6 out of 8 predictions [Fig\ref{fig3}(a), (b)]. For example, when we take the products of the equations (\ref{AVN1}), (\ref{AVN3}), (\ref{AVN4}), and (\ref{AVN5}), each value $X_{\rm{i}}$ ($Y_{\rm{i}}$) appears twice while the product on the right is -1, which causes the AVN contradiction. 
For the fact that we cannot get perfect 1 or -1 due to experimental noise, a certain bound, error rate $\epsilon$, is introduced to identify the nonlocality \cite{ryff1997bell,PhysRevLett.115.260402}. The error rate $\epsilon$, which describes the degree to which we get the wrong events, gives quantitative estimates of the inconsistency between quantum mechanics and local realism.
To get the AVN contradiction, the error rate $\epsilon$ needs to be smaller than 1/4 (the bound for 4-photon and 3-photon GHZ states). 
Experimentally, the outcomes of measurement settings in Eqs.(\ref{AVN1}-\ref{AVN8}) are shown in Fig.\ref{fig3}(c) with the average error rate $\bar{\epsilon}= 0.191\pm0.021$ and the largest error rate $\epsilon_{\rm{max}}=0.218\pm0.021$, revealing the violation of local realism.
Error rate for 3-photon GHZ states is also below the bound of 1/4, showing in Tab.\ref{tab1}.

Mermin inequality is another witness for GHZ nonlocality, using the expression below,
\begin{eqnarray}
    M_4&=|XXXX-(XXXY+XXYX+XYXX\nonumber\\
    &+YXXX)-(XXYY+XYXY+XYYX\nonumber\\
    &+YXXY+YXYX+YYXX)+(XYYY\nonumber\\
    &+YXYY+YYXY+YYYX)+YYYY|.
\end{eqnarray}
For four-photon GHZ states, quantum mechanics predicts the maximum possible value of $8\sqrt{2}$, while the bound of local realism is 4. Note that the maximum violation is reached when the GHZ state is rotated with a phase and into the form: $\left|0000\right>+\textbf{e}^{i3\pi/4}\left|1111\right>$. 
The measured Mermin value of our experiment is $\left<M_4\right>=6.98\pm0.16$, showing a violation of local realism by 18.7 standard deviations. Projecting one photon in mode $1/\sqrt{2}\left(\left|0\right>+\left|1\right>\right)$, we get 3-photon GHZ states and measure 3-photon Mermin inequalities by using $M_3=|XXX-(XYY+YXY+YYX)|$, where the quantum value is 4 while the classical bound is 2. The average result for 3-photon GHZ states is $\left<M_3\right>=2.50\pm0.13$ with 3.8 standard deviations violating the Mermin inequality [Tab.\ref{tab1}].

\begin{table}[tbp]
    \centering
    \begin{tabular}{cccc}
    \hline
    Photons &  Fidelity & AVN $\epsilon_{\rm{max}}$ & Mermin $\left<M_3\right>$\\
    \hline
    (B,C,D) & $0.753\pm0.008$ & $0.236\pm0.031$ & $2.43\pm0.12$\\
    (A,C,D) & $0.736\pm0.009$ & $0.191\pm0.031$ & $2.57\pm0.12$\\
    (A,B,D) & $0.744\pm0.007$ & $0.214\pm0.030$ & $2.57\pm0.11$\\
    (A,B,C) & $0.736\pm0.008$ & $0.234\pm0.029$ & $2.43\pm0.11$\\
    \hline
    \end{tabular}
    \caption{Results of three-photon GHZ states when projecting different photon. In three-photon case, the error rate $\epsilon$ smaller than 1/4, or the Mermin value $\left<M_3\right>$ higher than 2, reveals the violation of local realism and shows the quantum nonlocality.}
    \label{tab1}
\end{table}

\section{Discussion and Conclusion}
In this work, we have presented a silicon photonic chip for generating, manipulating, and analyzing multi-photon GHZ states. Advanced micro-ring sources provide high purity of photons without extra spectral filtering. Thus, high visibilities of path-correlation and heralded-HOM interference fringes are obtained. To witness the quantum nonlocality, we have shown the first AVN test and the Mermin inequality on an integrated photonic chip with both four- and three-photon GHZ states, demonstrating the reliability of current photonic integration techniques for testing fundamental properties of quantum systems. In contrast to the previous demonstrations of AVN and Mermin tests which are mainly based on photons generated from nonlinear bulk crystals \cite{pan2000experimental,PhysRevLett.91.180401,PhysRevLett.115.260402,erhard2018experimental}, the same spatial mode of single-mode waveguides and good interference easily-performed on MMIs guarantee high quality of path-entangled photons on chip. It makes path-mode encoding be an ideal degree of freedom which can be easily extended to high dimensions \cite{wang2018multidimensional,zhang2022resource,zheng2023multichip}. Besides, chip-based GHZ-entanglement devices with photons at telecom band can easily interface with fiber quantum network, showing potential for applications in multipartite quantum communication, such as quantum cryptographic conferencing \cite{PhysRevA.57.822,chen2006Multi,Murta2020qua,proietti2021exp,pickston2023conference} and quantum secret sharing \cite{PhysRevA.59.1829,PhysRevA.68.032309,chen2005experimental,bell2014experimental,lee2020quantum}.  Our work paves the way for employing integrated quantum devices to exploit and certify the underlying physics of complex entangled states and local realism, providing the potential for generating multi-photon high-dimension entangled states and for advanced quantum information processing.

\section*{Appendix A. On-chip path-mode parity sorter (PPS) and GHZ-entanglement generation via post-selection}
The controllable PPS consists of a MZI and two waveguides. It is a four-port device for input photon B and C with two path modes each. When the MZI in the middle acts as a waveguide crossing, photons in modes $\left|1_{\rm{B}}\right>$ and $\left|1_{\rm{C}}\right>$ interchange while modes $\left|0_{\rm{B}}\right>$ and $\left|0_{\rm{C}}\right>$ remain. Note that $\left|0_{\rm{B}}\right>$ ($\left|0_{\rm{C}}\right>$) stands for the photon in waveguide path B0 (C0) shown in Fig.\ref{fig1}. Thus, the coincidence detection between the PPS outputs B and C implies that two photons B and C are both in mode |0⟩ or both in mode |1⟩. In the experiment of four-photon generation, there are possibilities of two pairs emitting from one source module (S1 or S2), which would give two photons at the same inputs into the PPS. So the detection of photons in A and D is important, which heralds each source module emits one pair of entangled photons via post-selection. As the trigger photon has the same path mode as the heralded photon (Eq.\ref{eq1}), the four-fold coincidence of photons in outputs A-D implies four photons are all in mode |0⟩ or all in mode |1⟩, resulting in a post-selected four-photon GHZ state.

\section*{Appendix B. Relationship between fidelity and visibility of two-qubit entanglement}
Uncorrelated noise, from multi-pair emission or photons’ distinguishability, limits the quality of the produced two-photon Bell states. In another word, these noises don’t give the quantum interference with the weight equal to $(1-V)$. Hence, the quantum state can be written in the form of a Werner state,
\begin{equation*}
    \rho=V\left|\Psi_2\right>\left<\Psi_2\right|+(1-V)\mathbbm{1}/4,
\end{equation*}
where we assume the introduced noise is a white noise $\mathbbm{1}/4$. So we can calculate the state fidelity to the Bell state $\left|\Psi_2\right>$,
\begin{equation*}
    F=V+(1-V)/4=(1+3V)/4.
\end{equation*}

\section*{Appendix C. Visibility of the heralded-HOM fringes scanning the splitting ratio of MZI}
In the heralded-HOM interference, idler photons are used as triggers while signal photons are mixed to perform HOM interference. The spectral distinguishability and the impurity of signal photons limit the visibility of the heralded-HOM fringe. With the triggers of idler photons A and D, the heralded photon state can be written as, $B_{R2}^+ C_{R3}^+\left|vac.\right>$. Assuming that purity of signal photons is the same, we have $B_{R2}^+=\Sigma\lambda_i(b_{i,R2})^+$ and $C_{R3}^+=\Sigma\lambda_j (c_{j,R3})^+$, where $B^+(b^+ )$, $C^+(c^+)$ denote the inputs (and outputs) of the MZI; the subscript $R2$ or $R3$ shows the origin of photons. Then, the purity is defined as $Pur.=\Sigma\lambda_i^4$, and the spectral overlap is quantified with the correlation function $\sigma=\left<b_{i,R2}(b_{i,R3})^+\right>=\left<c_{i,R2}(c_{i,R3})^+\right>$. 
When the MZI acts as passing (or crossing) waveguides to identity (or swap) two photons B and C, the coincidence gives the maximum of the fringe $CC_{max}\sim1$, which is twice the maximum of the flat portion of HOM interference scanning the photon delay. 
When the MZI acts as a 50:50 BS, HOM interference is performed with the bunched photons at the outputs. The post-selected photon state, $1/2\left(B_{R2}^+ C_{R3}^+-B_{R3}^+ C_{R2}^+ \right)\left|vac.\right>$, gives the minimum coincidence of the fringe $CC_{min}\sim\left(1-\sigma^2Pur.\right)/2$. So, the visibility of heralded-HOM fringes scanning the splitting ratio of MZI is,
\begin{equation*}
    V_{HHOM}=\left(CC_{max}/2-CC_{min}\right)/\left(CC_{max}/2\right)=\sigma^2Pur.
\end{equation*}

\section*{Acknowledgments}
    This research was supported by the National Key Research and Development Program of China (Grants Nos. 2022YFE0137000, 2019YFA0308704), the Leading-Edge Technology Program of Jiangsu Natural Science Foundation (Grant No. BK20192001), the Fundamental Research Funds for the Central Universities, and the Innovation Program for Quantum Science and Technology (Grants Nos. 2021ZD0300700 and 2021ZD0301500).

\section*{Disclosures}
The authors declare that there are no conflicts of interest.

\bibliography{GHZ_bib}

\end{document}